\documentclass[aps,pra,twocolumn,showpacs,groupedaddress]{revtex4}
\usepackage{graphicx}

\begin{document}


\title{Optimal conversion of Bose-Einstein condensed atoms into molecules via a Feshbach resonance}

\author{Jaeyoon Jeong, Chris P. Search, and Ivana Djuric}
\affiliation{Department of Physics and Engineering Physics,
Stevens Institute of Technology, Hoboken, NJ 07030}

\date{\today}

\begin{abstract}
In many experiments involving conversion of quantum degenerate
atomic gases into molecular dimers via a Feshbach resonance, an
external magnetic field is linearly swept from above the resonance
to below resonance. In the adiabatic limit, the fraction of atoms
converted into molecules is independent of the functional form of
the sweep and is predicted to be $100\%$. However, for
non-adiabatic sweeps through resonance, Landau-Zener theory
predicts that a linear sweep will result in a negligible
production of molecules. Here we employ a genetic algorithm to
determine the functional time dependence of the magnetic field
that produces the maximum number of molecules for sweep times that
are comparable to the period of resonant atom-molecule
oscillations, $2\pi\Omega_{Rabi}^{-1}$. The optimal sweep through
resonance indicates that more than $95\%$ of the atoms can be
converted into molecules for sweep times as short as
$2\pi\Omega_{Rabi}^{-1}$ while the linear sweep results in a
conversion of only a few percent. We also find that the
qualitative form of the optimal sweep is independent of the
strength of the two-body interactions between atoms and molecules
and the width of the resonance.
\end{abstract}

 \maketitle

\section{Introduction}
Unlike alkali atoms, molecules can not be directly cooled using
the laser cooling techniques that led to Bose-Einstein
condensation because of the complex rotational-vibrational
spectrum of the molecules. As a result, two-photon Raman
photoassociation and Feshbach resonances have become the standard
tools to create translationally cold molecules starting from
ultra-cold atomic gases. The conversion of a macroscopic number of
quantum degenerate atoms into molecular dimers starting from
either a Bose-Einstein condensate \cite{mol-1,durr,xu-2003,herbig}
or a Fermi gas
\cite{regal-2003,strecker-2003,jochim-2003,cubizolles-2003} has
been observed by several experimental groups using a Feshbach
resonance. This work culminated in the formation of a molecular
Bose-Einstein condensate (MBEC) \cite{mol-BEC-K,mol-BEC-Li}.

Until recently the field of molecular optics was in the same state
as atom optics before BEC. Diffraction and interferometry had been
demonstrated \cite{mol-diffract} but there was no source of high
density phase coherent monoenergetic molecules analogous to a
laser that could be used to observe nonlinear and quantum optical
effects. Indeed, the ability to now coherently produce molecules
with a high phase space density opens up new avenues of research
in the area of matter-wave optics such as lasing and matter wave
amplifications with molecular fields, nonlinear mixing of atomic
and molecular matter waves, and the generation of entangled atoms
by `down conversion' of the molecules. Recent experiment have
shown the phase coherent and momentum conserving nature of matter
wave second harmonic generation \cite{molecule-optics}.

From the perspective of atom optics, the conversion of atoms into
molecules via a Feshbach resonance or photo-association is the
matter-wave analog of second harmonic generation of photons in a
nonlinear crystal with a $\chi^{(2)}$ susceptibility, which has
been used to create entangled photon states. From another
perspective, these methods can be viewed as the first step towards
`quantum super-chemistry' where chemical reactions occur in a
controllable phase coherent manner and exhibit novel quantum
features such as interference and bosonic
stimulation\cite{heinzen}. Also, the ability to create molecules
with permanent dipole moments using a heteronuclear Feshbach
resonance \cite{hetero-feshbach} opens up the possibility of
studying bosonic systems with anisotropic interactions
\cite{dipole}. Dipolar molecules can be stored in electrostatic
storage rings \cite{Crompvoets} that have an area that is twenty
five times larger than the largest neutral atom magnetic storage
ring \cite{arnold}. Storage rings for dipolar molecules therefore
have the potential to be used for high precision rotation sensors
based on the Sagnac effect, which is proportional to area enclosed
by the ring.

Whatever the reason for creating ultracold molecules, it is highly
desirable to have an efficient means of production. While the
molecules created via a Feshbach resonance are translationally
very cold, they are vibrationally very hot and can decay to lower
lying vibrational states via exoergic inelastic collisions with
atoms or other molecules. For an atomic BEC, the molecular
two-body decay rates are of the order $10^{-11}-10^{-10}cm^3/s$,
which gives a lifetime of $100\mu s$ for a typical atomic density
\cite{xu-2003,yurovsky,mukaiyama}. It is therefore important to be
able to create the molecules as quickly as possible so that
experiments can be performed on them before they are lost from the
system. (In this paper, we will not consider molecules created
from an atomic Fermi gas, which can have lifetimes on the order of
a second due to Pauli blocking of collisions \cite{petrov}.)

In experiments using Feshbach resonances in atomic BECs, adiabatic
rapid passage is used to convert atoms into molecules by 'slowly'
varying an external magnetic field from above resonance, where the
energy of the molecular state lies above that of the atoms to
below resonance, where molecules are the stable ground state. When
the magnetic field strength is swept linearly across the Feshbach
resonance, the atomic and molecular populations can be predicted
by the Landau-Zener (LZ) formulae for a two-level quantum system
\cite{bates}. If there are initially no molecules at $t\rightarrow
-\infty$, the probability at $t\rightarrow +\infty$ of a pair of
atoms remaining unconverted, $P_{atom}$, and of being converted
into molecules, $P_{molecule}$, is given by \cite{bates},
\begin{eqnarray}
P_{atoms}&=&e^{-2\pi \delta_{LZ}} \\
P_{molecules}&=&1-e^{-2\pi \delta_{LZ}} \label{P_mol_LZ} \\
\delta_{LZ}&=&\Omega_{Rabi}^2/4|\dot{\Delta}|
\end{eqnarray}
where $\Omega_{Rabi}$ is the Rabi frequency coupling the two
states and $\Delta(t)=|\dot{\Delta}|t$ is the energy difference
(detuning) between the two states. The Landau-Zener (LZ) factor
$\delta_{LZ}$ is the main parameter that describes the
atom-molecule conversion in theoretical models \cite{goral}.

In the absence of decay of the atoms and molecules, $100\%$
molecular conversion is expected from the LZ theory for
sufficiently small $|\dot{\Delta}|$. If the detuning varies slowly
enough to satisfy the adiabatic condition, $\delta_{LZ}> 1$, the
initial atomic BEC will adiabatically evolve into a molecular BEC.
In reality, the conversation rates are always much smaller than
that predicted by LZ theory due to molecular vibrational decay
\cite{yurovsky,mukaiyama}. Experiments with atomic BECs have
yielded molecular conversion efficiencies of only $\sim 5-10\%$
with a large fraction of the atoms being lost during the sweep
\cite{durr,herbig}. The missing fraction of atoms in the
experiments are attributed to vibrational decay of the molecules,
although spontaneous dissociation of molecules caused by inelastic
spin flips has been shown to be important in $^{85}Rb$
\cite{thompson}. This was recently confirmed by experiments in an
optical lattice where $^{87}Rb$ dimers were created with $95\%$
efficiency on lattice sites initially containing only two atoms so
that vibrational decay of the resulting molecule was impossible
\cite{thalhammer}.

Motivated by the need to create molecules faster than atom and
molecule loss rates, we have explored atom-molecule conversion via
a Feshbach resonance in the non-adiabatic regime. The sweep times
that we are primarily interested in are shorter than the molecular
lifetimes and therefore we ignore decay processes. To this end we
have employed a genetic algorithm to determine the magnetic field
sweep that maximizes the fraction of atoms converted into
molecules for sweeps as short as $2\pi\Omega_{Rabi}^{-1}$. We find
that the optimal sweep shows a conversion efficiency in excess of
$95\%$ when the linear sweep results in a conversion of only a few
percent. Moreover, the general form of the optimal sweep is
independent of the strength of the resonance and the strength of
the two-body interactions between atoms and molecules. These
results indicate that higher conversion efficiencies could be
obtained in experiments with the added bonus of longer times to
perform experiments on the molecules. A recent experiment on a
$Cs$ atomic BEC showed that up to $30\%$ of the atoms could be
converted into molecules using a sweep through resonance that is
similar to the one we propose here \cite{mark}. However, the
duration of their sweep is much longer than what we consider here
and can be considered adiabatic.

This paper is organized as follows. In section II we explain our
physical model and in section III we discuss the genetic algorithm
used. In section IV we compare the results obtained from the
genetic algorithm with a linear LZ sweep through resonance.
Section V presents some conclusions and  future directions for
this work.

\section{Physical Model}
A Feshbach resonance is a collisional resonance between a pair of
free colliding atoms and a molecular bound state of those atoms.
The coupling between the atomic and molecular states is due to the
hyperfine interaction, which can result in spin flip of the
electrons \cite{timmermans}. The colliding atom state is usually
referred to as the open channel while the molecular state is known
as the closed channel since it can only be accessed by a spin flip
of one of the atoms. Because the atomic and molecular states have
different magnetic moments, the energy difference of the atoms and
moleculs can be tuned using an external magnetic field.

At zero temperature, a weakly interacting BEC of atoms can be
described by the semi-classical Gross-Pitaevskii equation. These
equations can readily be extended to include a molecular BEC
produced by coherent interconversion of atoms into molecules via a
Feshbach resonance \cite{timmermans, goral, kohler},
\begin{widetext}
\begin{eqnarray}
i\hbar \dot{\phi}_1&=&\left(\frac{\hbar^2}{2m_1}\nabla^2+ V_1({\bf
x})+U_{11}|\phi_1|^2+U_{12}|\phi_2|^2\right)\phi_1
+\chi\phi_1^{*}\phi_2 \label{GP-1}
\\
i\hbar
\dot{\phi}_2&=&\left(-\hbar\Delta(t)+\frac{\hbar^2}{2m_2}\nabla^2+
V_2({\bf x})+U_{22}|\phi_2|^2+U_{12}|\phi_1|^2\right)\phi_2
+\chi\phi_1^2/2 \label{GP-2}
\end{eqnarray}
\end{widetext}
where $\phi_1$ is the atomic BEC wave function and $\phi_2$ that
of the molecules. Here, $m_i$ are the masses and $V_i(x)$ the
external trapping potential of the atoms ($i=1$) and molecules
($i=2$). The two-body interaction is given by a contact potential,
$V({\bf r}-{\bf r'})=U_{ij}\delta({\bf r}-{\bf r'})$, with the
coupling constants $U_{ij}=2\pi\hbar^2a_{ij}/m_{ij}$ where
$a_{11}$ is the atom-atom s-wave scattering length, $a_{22}$ the
molecule-molecule scattering length, $a_{12}$ the atom-molecule
scattering length, and $m_{ij}=m_im_j/(m_i+m_j)$. $\chi$ is the
coupling between the open channel and the closed channel defined
by $\chi=[2\pi\hbar^2a_{11}\Delta\mu\Delta B/m]^{1/2}$, where
$\Delta \mu$ is difference between magnetic moments of an atomic
pair and a molecule and $\Delta B$ is the width of resonance
\cite{timmermans}.

$\hbar\Delta(t)=\Delta\mu (B(t)-B_0)$ is the Zeeman energy
difference between the molecules and atoms such that $B_0$ is the
magnetic field strength at which the energy of the molecule equals
the open channel collision threshold. For $\Delta(t)<0$ the atoms
are the lowest energy state while for $\Delta(t)>0$, the molecular
state is the lowest energy state of the system. Although we ignore
vibrational decay of the molecules in our calculations, it can be
easily included by making the substitution $\Delta(t)\rightarrow
\Delta(t)-i\kappa_a |\phi_1(t)|^2-i\kappa_m|\phi_2(t)|^2$
\cite{timmermans}.

Unless stated otherwise we used the following values for the
two-body interactions, $U_{11}/\hbar = 4.96\times 10^{-17} m^3/s$,
$U_{12}/\hbar = -6.44\times 10^{-17} m^3/s$, $U_{22}/\hbar =
2.48\times 10^{-17} m^3/s$ corresponding to the scattering lengths
of an $^{87}Rb$ condensate. For the atom-molecule coupling constant
we use $\chi/\hbar = 1.91\times 10^{-5} m^{3/2}/s$ in all of our
simulations \cite{goral}. We also take $V({\bf x})=0$, which implies
that $\phi_1$ and $\phi_2$ will be spatially homogenous. This is
justified even for trapped gases in the local density approximation
\cite{dalfovo}.

In this paper, we are interested in the functional form of
$\Delta(t)$ that optimizes the conversion of atoms into molecules
starting from the initial condition $\phi_1=\sqrt{n_0}$,
$\phi_2=0$, and $\Delta(t)<0$. We also impose the constraint that
the initial, $\Delta(-T/2)$, and final, $\Delta(T/2)$, values of
the detuning, and the sweep time, $T$, are fixed. We employ a
genetic algorithm to find the optimal sweep for range of different
$T$. The conversion efficiency of the sweep is given by the
probability that a pair of atoms becomes a molecule,
$P_{molecule}=2|\phi_2(T/2)|^2/n_0$.

\section{Genetic Algorithm}
Genetic algorithms (GAs) have become a widely used tool for
solving optimization problems that depend on a large number of
variables \cite{coley}. These multidimensional optimization
techniques proceed by parameterizing an objective function in
terms of a finite set of coefficients, the chromosome. The genetic
algorithm operates on a set of chromosomes, which represent the
members of a population.

The goal is to find the chromosome that is the global minimum of the
objective function. In order to find the optimal chromosome, GAs
employ Darwin's principle of survival of the fittest. At each
generation, the chromosomes in the population compete with each
other for survival. The members of the population that do the best
job of minimizing the objective function, survive to the next
generation while those that do a poor job are eliminated. The
surviving members are allowed to produce offspring for the next
generation by some prescribed mating rules that mix elements from
their chromosomes (crossover). Random mutations are also introduced
into the offspring chromosomes at each generation. This process
repeats itself until the change in the optimum solution between
successive generations is less than some threshold value.

We want to optimize the molecular fraction obtained by sweeping
$\Delta(t)$ from $\Delta(-T/2)=-|\Delta_0|$ to
$\Delta(T/2)=|\Delta_0|$. The detuning is parameterized in terms
of either a power series,
\begin{equation}
\Delta(t)=\sum^{N}_{n=1} C_n t^n \label{detuning-1}
\end{equation}
or a Fourier series,
\begin{equation}
\Delta(t)=\sum^N_{n=1} B_n\sin\left(\frac{n\pi t}{T}\right).
\label{detuning-2}
\end{equation}
The coefficients $C_n$ or $B_n$ are the elements of a chromosome,
where we used $N=160$ coefficients for Eq. (\ref{detuning-1}) and
$N=320$ coefficients for Eq. (\ref{detuning-2}). The initial
population consists of $N_{pop}=20$ randomly chosen chromosomes.
We solve the coupled Gross-Pitaevskii equations, Eqs. \ref{GP-1}
and \ref{GP-2}, for each detuning function in the current
generation. The half of the population that produces the largest
fraction of molecules are kept and allowed to breed to produce
$N_{pop}/2$ new chromosomes while the other half are discarded.

Crossover is accomplished by choosing pairs of chromosomes and
representing them as binary strings. A random point in the strings
is chosen and all information to the right of that point is
swapped. The mutation operator randomly flips the value of single
bits within the binary strings of the offspring chromosome. This
process of selection, crossover, and mutation continues until the
difference in optimal conversion efficiency between $50$
successive generations is less than $0.1\%$. This stopping
condition is because the GA can become stuck for several
generations at local minima of the objective function.

\section{Numerical Results}
The conversion efficiencies of atoms into molecules determined by
the GA are characterized by the initial and final values of
detuning functions ($|\Delta_0|$), sweep time (T), and the mean
field energy shifts for the atoms and molecules, which are
proportional to $U_{11}$, $U_{12}$, and $U_{22}$. In order to make
our results independent of the strength of the resonance, all
times and frequencies are scaled relative to the initial Rabi
frequency,
$T_{Rabi}/2\pi=\Omega_{Rabi}^{-1}=\hbar/\chi\sqrt{n_0}$. However,
to provide a feeling for the relevant time scales, we note that
$T_{Rabi}=16\mu s$ for the initial atomic density of $n_0=4\times
10^{14}cm^{-3}$ that we use in our simulations.

We first investigated the molecular conversion efficiency as a
function of magnetic field sweep time with
$|\Delta_0|=50\Omega_{Rabi}$. This initial value of the detuning
is large enough to decouple the atoms and molecules. The circles
and squares in Fig. 1(a) indicate the conversion efficiency for a
linear sweep and optimal sweep, respectively. For long enough
sweep times, $\tau\equiv T/T_{Rabi}> 100$, both sweeps show very
large conversion efficiencies of around $95\%$ or more. In this
limit, the sweeps are adiabatic and the conversion efficiency is
therefore independent of the shape of the detuning function.

As $\tau$ is decreased, the optimal sweep is still able to convert
almost all atoms into molecules while the conversion efficiency of
a linear sweep decreases rapidly. For $\tau\sim 1$, the conversion
efficiency of a linear sweep is only a few percent. This is easily
understood by noting that as $\tau$ decreases the linear sweep is
no longer adiabatic. For $\tau=100$, $\delta_{LZ}=1.57$ and
according to Eq. (\ref{P_mol_LZ}), $P_{molecule}=0.99$ while for
$\tau=1$ the LZ factor decreases to $\delta_{LZ}=0.015$ giving a
molecular fraction of only $P_{molecule}=0.089$. These values
agree qualitatively with the numerical solution of the GP
equations in Fig. 1(a), which indicates that the LZ factor is the
main parameter controlling the atom-molecule conversion. By
contrast, the optimal sweeps show excellent conversion even for
very small sweep times. For comparison the optimal sweep gives
$98.48\%$ and $99.82\%$ conversion for $\tau=1$ and $2$, while the
linear sweep gives $6.74\%$ and $10.64\%$, respectively.

\begin{figure}
\includegraphics*[width=8cm,height=8cm]{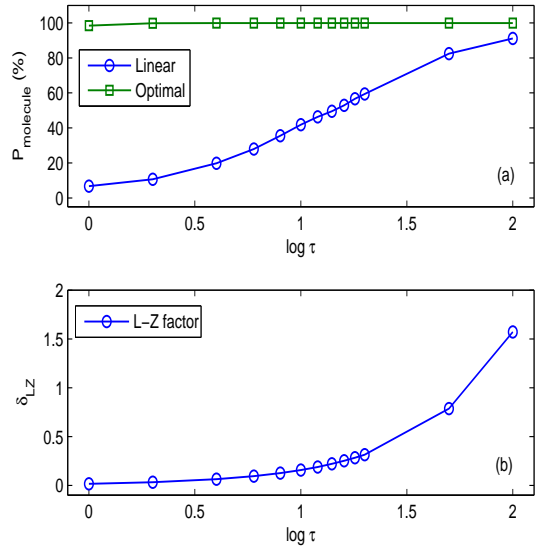}
\caption{(Color online)(a) Fraction of atoms converted into
molecules as a function of $\log_{10}\tau$ where $\tau$ is the
dimensionless magnetic field sweep time, $\tau=T/T_{Rabi}$ for
fixed initial and final detunings $\Delta (-\tau/2)=-\Delta
(\tau/2)=-|\Delta_0|=-50 \Omega_{Rabi}$. Here we chose
$T_{Rabi}=1.6465\times 10^{-5}s$. The circles and squares indicate
the conversion rate for the linear and optimal sweeps,
respectively. (b) Landau-Zener factors, $\delta_{LZ}$, for linear
sweep.}
\end{figure}

Fig. 2 shows the optimal sweep for $\tau=2$ using both Eq.
(\ref{detuning-1}) and (\ref{detuning-2}). The power series and
the Fourier series result are almost identical except for the
small noise fluctuations present in the Fourier series. The noise
in the Fourier series GA varied from run to run but always had an
average amplitude less than $\Omega_{Rabi}$. The noise was the
result of the high frequency terms in the Fourier series, $n\pi/T
\gg \Omega_{Rabi}, |\Delta_0|$, which oscillate much faster than
all other time scales in the Gross-Pitaevskii equations. We
therefore do not expect these fluctuations to have any appreciable
effect on the dynamics. These high frequency terms needed to be
included in $\Delta(t)$ to produce the large slope at the
beginning and end of the sweep.

To study the effect of these fluctuations, we superimposed
randomly generated noise on top of the optimal sweep obtained from
the power series, Eq. (\ref{detuning-1}). The average amplitude of
the noise was varied between $0.01\Omega_{Rabi}$ and
$100\Omega_{Rabi}$. We then calculated the conversion efficiency
of the power series detuning with the added noise. For noise
amplitudes less than $\Omega_{Rabi}$ there was no noticeable
change in the conversion efficiency or the dynamics of the
atom-molecule conversion. While noise amplitudes larger the
$\Omega_{Rabi}$ significantly effect the conversion dynamics and
efficiency. This indicates that as long as the size of the
fluctuations are small enough to not shift the two states into or
out of resonance with each other, there will be no observable
effect.

\begin{figure}
\includegraphics*[width=8cm,height=8cm]{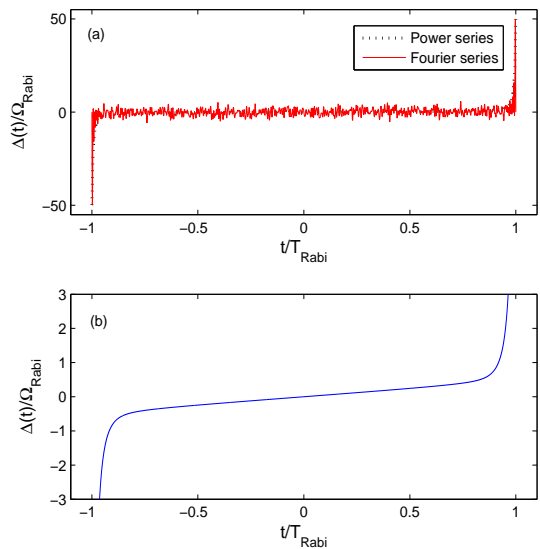}
\caption{(Color online) (a) The optimal sweep for $\tau=2$. The
dotted line is the optimal sweep obtained from Eq.
(\ref{detuning-1}) while the solid line is that obtained from Eq.
(\ref{detuning-2}). (b) Closeup of the optimal sweep obtained from
Eq. (\ref{detuning-1}) for the power series.}
\end{figure}

In order to understand why the optimal sweep is able to achieve
such high conversion efficiencies even for very small $\tau$, we
plot in Fig. 3 the {\em instantaneous} Landau-Zener factor
$\delta_{LZ}(t)=\Omega_{Rabi}^2(t)/4|\dot{\Delta}(t)|$ where
$\Omega_{Rabi}(t)=\chi|\phi_1(t)|$ and $\Delta(t)$ is the GA
optimal sweep. At the beginning and end of the sweep
$\delta_{LZ}(t)\ll 1$, which indicates that the sweep is
non-adiabatic. However, at these times the detuning is so large
that the atoms and molecules are decoupled. It is only for
$\Delta(t) < \Omega_{Rabi}(t)$ that the atom-molecule system is
resonant and atoms can be converted into molecules. At these times
$\delta_{LZ}(t)>1$, which implies that the system is actually
adiabatic .

As one can see from Fig. 2, the optimal sweep rapidly goes from
far off resonance to close to resonance and once near resonance
the sweep rate slows down to maintain adiabaticity. By using the
narrow time interval at the beginning and end of the sweep where
the detuning function changes the fastest, $\Delta t$, one can
estimate the frequency bandwidth of the sweep, $\Delta
E/\hbar\approx 1/\Delta t$, to be $\simeq 4.5 MHz$. This is
several orders of magnitude smaller than the typical energy
spacing between molecular vibrational levels in the closed
channel. Consequently, the coupling between atoms and more than
one molecular state can be ignored.

At the same time the gradient of the detuning is large enough to
prevent nonadiabatic oscillations, as one can see from Fig. 4(a).
On the other hand the linear sweep results in nonadiabatic
oscillations close to resonance that are the result of the
coupling between the dressed states that is proportional to
$\dot{\Delta}(t)/\Omega_{Rabi}$. In the optimal sweep the
atom-molecule conversion occurs gradually over almost the entire
sweep, while the conversion is limited to a very small window of
approximately $t\approx 0$ to $t\approx 0.1 \tau$ for the linear
sweep. The optimal sweep is similar to the switching scheme
demonstrated in \cite{mark} where the magnetic field was held at a
fixed value close to the resonance for a certain time ($t\geq
15ms$). The switching scheme produced more molecules than the
linear sweep because the atoms spent more time within the line
width of the resonance where molecules can be created. However, we
emphasize that their hold times near resonance were several orders
of magnitude longer than the nonadiabatic sweep times we consider
here.

To see whether the form of the optimal detuning was unique to Eqs.
(\ref{GP-1}) and (\ref{GP-2}) or were universal to two-state
systems, we tested our GA on a two-level quantum mechanical system
with the Hamiltonian,
\begin{equation}
H= \hbar \left( \begin{array}{cc}
  0 &  \Omega_{Rabi} \\
   \Omega_{Rabi} & -\Delta(t) \\
\end{array} \right)
\end{equation}
Fig. 5 shows the populations of the two levels $|a\rangle$ and
$|m\rangle$ with all of the population initially in $|a\rangle$.
The dynamics of the conversion of $|a\rangle$ into $|m\rangle$ is
qualitatively the same as the atoms and molecules as is the shape
of the optimal detuning function. This indicates that the general
form of the optimal $\Delta(t)$ is generic to any two state
system.

The mean field energy shifts of the atoms ($U_{11}|\phi_1|^2$, and
$U_{12}|\phi_2|^2$) and molecules ($U_{22}|\phi_2|^2$ and
$U_{12}|\phi_1|^2$) modifies the energies of the atom and
molecules so that the difference in the chemical potentials
between the atoms and molecules, $\Delta
\mu=\mu_{molecule}-\mu_{atom}$, to zeroth order in $\chi$ is
\cite{comment},
\begin{equation}
\Delta
\mu=-\hbar\Delta(t)+U_{22}|\phi_2|^2-U_{11}|\phi_1|^2+U_{12}\left(|\phi_1|^2-|\phi_2|^2\right)
\end{equation}
Consequently, even when $\Delta(t)=0$ the energy difference
between the atoms and molecules is not zero.

\begin{figure}
\includegraphics*[width=8cm,height=6cm]{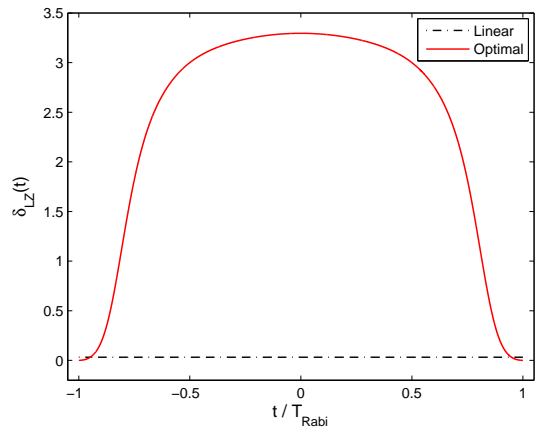}
\caption{(Color online) Instantaneous Landau-Zener factor,
$\delta_{LZ}(t)$ for the optimal sweep (solid line) and linear
sweep (dot-dashed line). The sweep time is $t=2T_{Rabi}$}
\end{figure}

\begin{figure}
\includegraphics*[width=8cm,height=8cm]{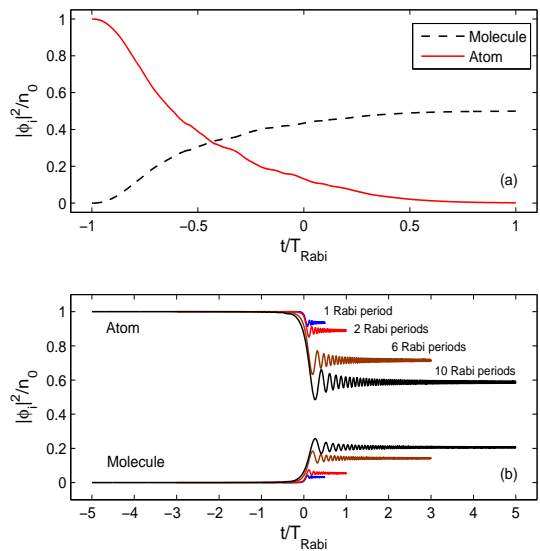}
\caption{(Color online) (a) Atomic, $|\phi_1|^2/n_0$, (solid line)
and molecular, $|\phi_2|^2/n_0$, (dashed line) fractions from the
optimal sweep for $\tau=2$. (b)Atomic and molecular fraction
obtained from linear sweep for $\tau=1,2,6,10$.}
\end{figure}

\begin{figure}
\includegraphics*[width=8cm,height=8cm]{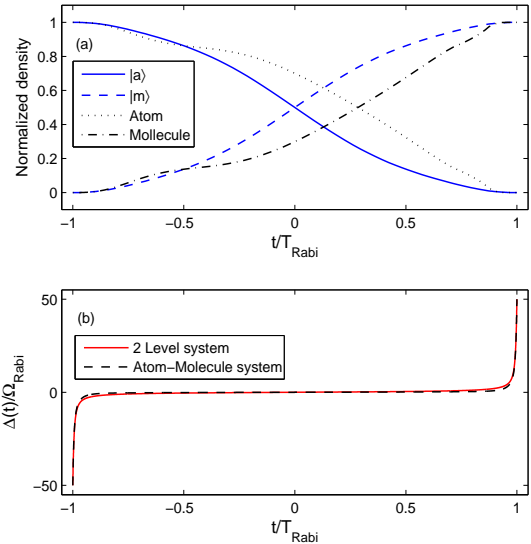}
\caption{(Color online)(a) Populations of two level quantum
system, $|a\rangle$ (solid line) and $|m\rangle$ (dashed line)
obtained from optimal sweep through resonance for $\tau=2$. Also
shown are atomic, $|\phi_1|^2/n_0$, (dotted) and molecular,
$2|\phi_2|^2/n_0$, (dashed dot) fractions for with
$U_{11}=U_{12}=U_{22}=0$ and initial Rabi frequency chosen to
match that of the two level system. (b)Optimal sweep, $\Delta(t)$,
obtained from Eq. \ref{detuning-1} for two level system (solid
line). For comparison, the optimal detuning for the atom-molecule
system is also shown (dashed line).}
\end{figure}

In order to test the effect of the mean field shifts on the
conversion process, we consider a range of possible values for the
coupling constants $U_{ij}$ to see how sensitive the shape of the
optimal sweep and the conversion efficiency is to variations of
these parameters. We took the coupling constants to be either
$0.1$, $1$, or $10$ times bigger than the values given in Sec. II,
which we denote here by $U_{ij}^{(0)}$. The maximum conversion
efficiency for each of the $27$ different combination of values
are shown in Fig. 6. The numbers in parentheses represent the
relative magnitudes of the three coupling constants. For example,
$(1,10,0.1)$ means that $U_{11}=U_{11}^{(0)}$,
$U_{12}=10U_{12}^{(0)}$, and $U_{22}=0.1U_{22}^{(0)}$.

In all cases, the optimal sweep shows very good conversion ($\geq
98\%$) compared to less than $5\%$ for the linear sweep even for
very short sweep times. For instance, with
$U_{11}=10U_{11}^{(0)}$, the optimal sweep still shows better
conversion than the linear sweep giving for example $\sim 99\%$
conversion compared to $<5\%$ for $\tau=2$. To understand this we
note that from Fig. 4(a), most of the conversion takes place at
the beginning of the sweep, $t/\tau\approx -1.0$ to $-0.5$ where
the atomic density is much larger than the molecular density. At
these times the molecular energy lies above that of the atoms,
$-\hbar \Delta(t)>0$ and the atomic mean field $U_{11}|\phi_1|^2$
shift reduces $\Delta \mu$.

As the $U_{ij}$ are varied, the qualitative form of the optimal
detuning are the same as Fig. 2. There are, however, quantitative
variations in the slope for different $U_{ij}$ as can be seen in
Fig. 7. We tested the optimal detuning for each of the the 27
combination of parameters denoted by $\Delta_{(l,m,n)}(t)$, on the
other 26 combination of parameters. In all cases the
$\Delta_{(l,m,n)}$ performed significantly better when applied to
the other combination of parameters than the linear sweep. These
results imply that the optimal detuning calculated for any set of
values of the two-body interaction can be used to convert atoms
into molecules with greater efficiency than a linear sweep.

\begin{figure}
\includegraphics*[width=8cm,height=8cm]{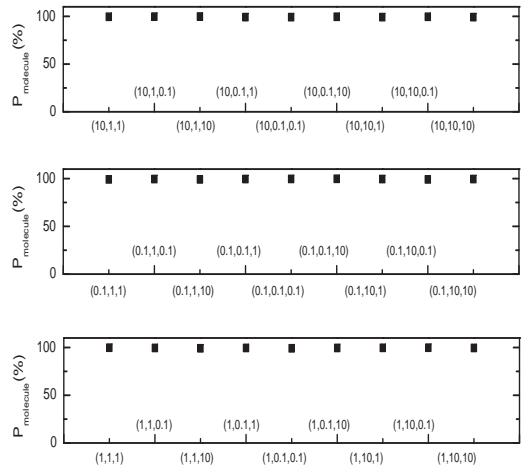}
\caption{Molecular fraction as a function of non-linear
interaction terms for $\tau=2$. The numbers in parentheses
indicate relative magnitude of the mean field coupling constants
$U_{11}$, $U_{12}$, and $U_{22}$, respectively. For instance, (10,
10, 0.1) means $U_{11}=10 U_{11}^{(0)}$, $U_{12}=10U_{12}^{(0)}$,
and $U_{22}=0.1 U_{22}^{(0)}$.}
\end{figure}

\begin{figure}
\includegraphics*[width=8cm,height=8cm]{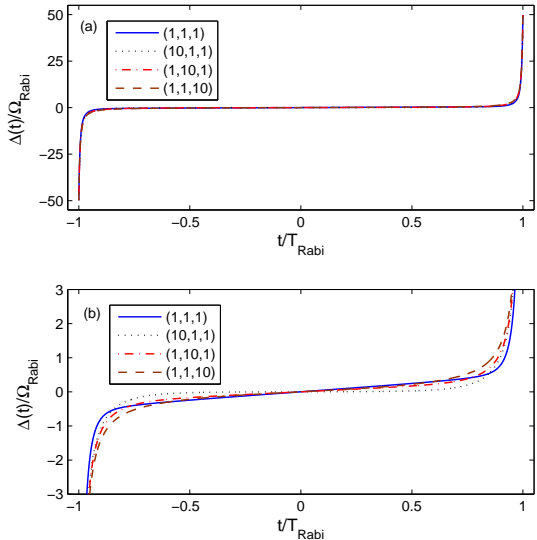}
\caption{(Color online) Optimal detuning function for several
different values of $U_{11}$, $U_{12}$, and $U_{22}$.}
\end{figure}

\section{Discussion}
Very close to resonance, mean field theory breaks down due to the
divergence of the scattering length and higher order quantum
correlations can become important \cite{kohler}. These higher
order correlations have been shown to be important for time
dependent experiments such as in Ref. \cite{mol-1} where the atoms
and molecules are placed in a coherent superposition state and
allowed to evolve freely \cite{goral-PRA}. However, theoretical
calculations that compare mean field theory and microscopic
quantum calculations that account for higher order correlations
have been shown to give nearly identical results for the molecular
conversion efficiency for downward magnetic field sweeps through a
resonance \cite{goral,kohler}. Although these calculations were
for linear sweeps, the agreement was very good for a range of
$\dot{\Delta}$ that included non-adiabatic sweeps for which
Landau-Zener theory gives $P_{molecules} \ll 1$. These results
support our use of mean field theory to model the atom-molecule
conversion. It would, however, be of great interest to extend the
work of Ref. \cite{goral} and compare mean field theory to quantum
calculations for nonlinear magnetic field sweeps.

Throughout this paper we have ignored the effect of the decay of
the molecules. If we assume a two-body inelastic decay rate of $5
\times 10^{-11}cm^3/s$ \cite{mukaiyama} for molecule-molecule and
atom-molecule collisions, we get a lifetime of $\tau_{loss}=50\mu
s$ for the initial atomic density of $4\times 10^{14}cm^3$.
Assuming a simple exponential decay of the molecule number, we can
estimate the fraction of molecules lost during the sweep by
$\exp(-T/\tau_{loss})$. This estimate actually overestimates the
molecule loss since it does not take into account the fact that
$\tau_{loss}$ is not constant but instead increases as the density
of the gas decreases with time. $\exp(-T/\tau_{loss})$ should,
nevertheless, provide a lower bound on the fraction of molecules
lost during a fast sweep. For $T=T_{Rabi}=16\mu s$, the molecular
conversion efficiency is reduced by a factor of
$\exp(-T_{Rabi}/\tau_{loss})=0.73$. Therefore, for a sweep time of
$T_{Rabi}$ the conversion efficiency using our optimal sweep and
taking into account losses will be approximately
$\exp(-T_{Rabi}/\tau_{loss})\times 98.48\%=72\%$.

In conclusion, we have applied a genetic algorithm to study the
optimal time dependence of an applied magnetic field, $B(t)$, to
convert atoms into molecules using a Feshbach resonance. Instead
of the conventional linear sweep, we propose a nonlinear sweep
through resonance that is characterized by a rapid approach of
$B(t)$ towards the resonance position followed by a gradual change
in $B(t)$ close to resonance and finally a rapid change in $B(t)$
at the end of the sweep away from the resonance. This nonlinear
sweep can result in almost $99\%$ conversion of atoms into
molecules even for sweep times as short as one Rabi period. We
have shown that the qualitative form of this nonlinear sweep is
independent of the specific nature of the two state system. We
plan to extend this work by using a genetic algorithm to optimize
molecule production via two-photon Raman photoassociation. In that
case both the laser frequencies and intensities of the two lasers
can be treated as time dependent quantities to be optimized.


\begin{references}
\bibitem{comment} In chemical equilibrium
$\mu_{molecule}=2\mu_{atom}$. However, we are not considering the
static equilibrium case.
\end{references}
\end{document}